\input harvmac
\def\half{{1 \over 2}}
\def\dzm{{\partial_z}}

\def\dg{{\dagger}}

\def\p{{\partial}}
\def\s{{\sigma}}

\def\L{{\Lambda}}

\def\a {{\alpha}}
\def\b {{\beta}}
\def\m {{\mu}}
\def\n {{\nu}}
\def\g {{\gamma}}
\def\d {{\delta}}

\def\ad {{\dot\alpha}}
\def\bd {{\dot\beta}}
\def\k {{\kappa}}

\def\t {{\theta}}

\def\tba {{\bar\theta^\ad}}

\def \ad {{\dot \a}}
\def \bd {{\dot \b}}
\def \t {{\theta}}
\def \tb {{\bar\theta}}

\def \Gtp {{\tilde G^+}}

\Title{\vbox{\hbox{IFUSP-P-1212}}}
{\vbox{\centerline{\bf A New Description of the
Superstring}}}
\bigskip\centerline{Nathan Berkovits}
\bigskip\centerline{Dept. de F\'{\i}sica Matem\'atica, Univ. de S\~ao Paulo}
\centerline{CP 20516, S\~ao Paulo, SP 01498, BRASIL}
\centerline{and}
\centerline{IMECC, Univ. de Campinas}
\centerline{CP 1170, Campinas, SP 13100, BRASIL}
\bigskip\centerline{e-mail: nberkovi@snfma2.if.usp.br}
\vskip .2in
This is a review of the new manifestly spacetime-supersymmetric description 
of the superstring.  The new description contains N=2 worldsheet 
supersymmetry, and is related by a field redefinition to the standard RNS 
description. It is especially convenient for four-dimensional 
compactifications since SO(3,1) super-Poincar\'e invariance can be made 
manifest. Parts of this work have been done in collaboration with
Warren Siegel and Cumrun Vafa.

This review is based on lectures given at the VIII J.A. Swieca summer
school and should be easily accesible to anyone familiar with the RNS
superstring description.

\Date{April 1996}
\newsec {Introduction}

This review is based on five lectures given at the VIII J.A.
Swieca summer school in Rio
de Janeiro in 1995. The first half of the lectures consisted of an 
introduction to string theory, while the second half concentrated
on the new spacetime-supersymmetric description of the superstring.
Since much better introductions to string theory are available, this
review will only describe the second half of the lectures.

There are two ways to look for a manifestly spacetime-supersymmetric
description 
of the superstring. One can either try to spacetime-supersymmetrize
the Ramond-Neveu-Schwarz
description, or one can try to covariantize the light-cone Green-Schwarz
description. As will be shown in this review, both of these approaches lead
to the same answer. Although historically, the GS approach was used earlier, 
I shall start with the RNS approach since it is familiar to a wider audience.

After reviewing the RNS superstring in section 2, I discuss the new
spacetime-supersymmetric description in section 3. This new description
has critical
N=2 worldsheet superconformal invariance and is related to the RNS description
by a field redefinition. For compactifications to four dimensions,
it allows the full SO(3,1) super-Poincar\'e invariance to be made
manifest.\ref\fourme{N. Berkovits, Nucl. Phys. B431 (1994) p.258.} 

In section 4, I introduce the N=4 topological method, which was
developed together with Vafa. This method can be
used to calculate scattering amplitudes for any string
with critical N=2 superconformal invariance, and is
much simpler than conventional N=2 techniques.
\ref\top{N. Berkovits and C. Vafa, Nucl. Phys. B433 (1995)
p.123.}.
In section 5, I discuss the relationship 
with conventional and twistor-like Green-Schwarz
descriptions, and in
section 6, I conclude with a list of applications for the new superstring
description.

\newsec {Review of the RNS Description}

\subsec {Worldsheet fields and  N=1 superconformal generators}

The standard RNS description of the superstring is a critical c=15
representation of N=1 worldsheet superconformal invariance.\ref\friedan
{D. Friedan, E. Martinec and S. Shenker, Nucl. Phys. B271 (1986) p.93.}
The simplest such
representation (corresponding to an uncompactified ten-dimensional manifold)
consists of ten bosonic worldsheet scalars, $x^\m$ ($\m$=0 to 9), 
and ten fermionic
worldsheet spinors, $\psi_{L,R}^\m$. ($\psi_L^\m$ and $\psi_R^\m$ are left
and right-moving components of the worldsheet spinor. When the
$L/R$ index is suppressed, we shall always mean the
left-moving component.
Although this review will only discuss the open and closed superstring, all
methods can easily be extended to the heterotic superstring.) 

The action and left-moving
c=15 N=1 superconformal generators for this representation
are 
\eqn\act{S={1\over 2\pi}\int d^2 z (\half \p x^\m \bar\p x_\m + 
i\psi_L^\m \bar\p\psi_{L\m}
+i\psi_R^\m\p\psi_{R\m})}
$$T=\half \p x^\m \p x_\m +{i\over 2} \psi^\m \p \psi_\m,\quad
G = \psi^\m \p x_\m$$
where $\p= {\p \over {\p z}}$.
and $\bar\p= {\p \over {\p\bar z}}$. 
Because the action is quadratic, the worlsheet fields have the free-field
OPE's:
\eqn\ope{
x^\m(y) x^\n(z) \to \eta^{\m\n}\log |y-z|^2,
\quad \psi^\m(y)\psi^\n(z) \to
i\eta^{\m\n} (y- z)^{-1}}
as $y \to z$ ($\eta^{\m\n}=(+,-,...,-)$). 
Using these OPE's, it is straightforward to show that
the above generators satisfy the c=15 N=1 super-Virasoro algebra:
\eqn\opea{
G(y) G(z)\to {{10i}\over{(y-z)^3}}+{{2iT}\over {y-z}},\quad
T(y)T(z) \to {{15}\over{2(y-z)^4}}+{{2T}\over{(y-z)^2}}
+{{\p T}\over {y-z}}.}

To gauge-fix the N=1 superconformal invariance, one needs a c=$-15$ 
ghost system consisting of a fermionic pair $(b,c)$ with weights 2 and $-1$,
and a bosonic pair $(\beta,\gamma)$ with weights ${3\over 2}$ 
and $-\half$. For constructing Ramond states and
spacetime-supersymmetry
generators, 
it is convenient to fermionize the bosonic
ghosts as
\eqn\gh{\beta = -i\p\xi e^{-\phi},\quad \gamma=i\eta e^{\phi}}
where $(\eta,\xi)$ are fermionic with weights (1,0), and $\phi$ is
a chiral boson with background charge $+2$. 
The OPE's for these fields are
\eqn\opec{b(y) c(z) \to i(y-z)^{-1},\quad
\xi(y) \eta(z) \to i(y-z)^{-1},\quad \phi(y) \phi(z) \to -\log (y-z)+{{i\pi}
\over 2}}
as $y\to z$.

Because the $\xi$ zero mode can not be expressed in terms of $\beta$
and $\gamma$, this fermionization procedure is only invertible on
the ``small'' Hilbert space, defined as the space of states $\Phi$
satisfying $[
\oint dz\eta, \Phi$]=0. The ``large'' Hilbert space, on the
other hand, is defined to also include states which are proportional to
the $\xi$ zero mode.

\subsec {Physical states}

A nilpotent BRST operator for this critical N=1 system is defined by
\eqn\Q{Q={1\over {2\pi}}
\oint dz ~[c(T+ib\p c+\half\p\phi\p\phi+\partial^2 \phi+i\eta\p\xi)+
i\eta e^\phi G +i b\eta\p\eta e^{2\phi} ]} 
where $T$ and $G$ are defined in \act.
Physical states are described by vertex operators $V$ which are in
the BRST cohomology of $Q$, i.e. $\{Q,V\}=0$ and $V \neq [Q,\Lambda]$
for any $\Lambda$ in the small Hilbert space. Note that if $\Lambda$ 
were allowed to live in the large Hilbert space, the cohomology would
be trivial since
$\{Q,V\}=0 $ implies $V=[Q, ic\xi\p\xi e^{-2\phi} V]$.  
Depending on the boundary conditions of $\psi^\m$,
the vertex operator can represent either a spacetime boson (Neveu-Schwarz
boundary conditions) or a spacetime fermion (Ramond boundary conditions).

Besides being in the BRST cohomology, a vertex operator must also satisfy
the following four conditions in order to uniquely represent a physical state:

1) The first condition is that the vertex operator is hermitian, i.e.
$V=V^\dg$. This type of reality condition is expected since
$V$ is interpreted as a second-quantized field, rather than
a first-quantized wave-function.

2) The second condition on the physical vertex operator is that it is
fermionic and
GSO projected, i.e. has no square-root cuts with the
spacetime-supersymmetry generator
\eqn\susy{q_a ={1\over {2\pi i}}\oint dz e^{{i\over 2}
(i\phi \pm \s_0 \pm \s_1\pm \s_2 \pm \s_3\pm \s_4)}}
where $\psi^9\pm\psi^0=e^{\pm i\s_0}$,
$\psi^j\pm i\psi^{j+4}=e^{\pm i\s_j}$ for $j$=1 to 4, and there are
an even number of $+$ signs in the exponent ($a$=1 to 16).
The spacetime-supersymmetry generator of \susy transforms Neveu-Schwarz
boundary conditions into Ramond boundary conditions, and 
this second condition 
ensures that the physical spectrum is spacetime-supersymmetric.

3) The third condition comes from the fact that
each physical state is represented by infinitely many vertex operators
in the BRST cohomology. This is because of the existence of the
picture-changing operator,
\eqn\pict{Z=\{Q,\xi\}= i e^{\phi} \psi_\m \p x^\m +ib\p\eta e^{2\phi}+i
\p(b\eta e^{2\phi} ) +i c\p\xi ,}
and the inverse picture-changing operator,
$$Y= i c \p\xi e^{-2\phi}.$$
Since $[Q,Z]=0$, $[Q,V]=0$ implies that $[Q,ZV]=0$. Also $ZV=\{Q,\Lambda\}$
implies that $V=\{Q, Y\Lambda\}$, so $ZV$ is in the BRST cohomology if
$V$ is. Similarly $Z^n V$ and $Y^n V$ are also in the BRST cohomology
for arbitrary positive integer $n$.

In order to choose a unique vertex operator for each physical state, it is
convenient to define the ``picture-counting'' operator
\eqn\picture{P={1\over {2\pi}}\oint dz (i\xi\eta -\p\phi).}
If $[P,V]= ipV$ where $p$ is the ``picture'' of $V$, then 
$[P,Z^n V]=i(p+n) V$ and 
$[P,Y^n V]=i(p-n) V$. One can therefore fix this overcounting by demanding
that $V$ sits in a certain picture (for example, one could demand that
all Neveu-Schwarz
vertex operators have picture $-1$ and all Ramond vertex operators
have picture $-\half$). However, since the spacetime-supersymmetry
generators of \susy carry picture,
such a restriction breaks manifest spacetime supersymmetry.

4) The fourth condition on the physical vertex operator is that it has
ghost-number $+1$. Although the ghost-number operator is usually defined by 
${1\over {2\pi}}
\oint dz (i bc + \b\g)={1\over {2\pi}}
\oint dz (i bc +\p\phi)$, we shall instead define it by
$$J_g={i\over {2\pi}}
\oint dz ( bc + \xi\eta)$$ so that ghost-number commutes with 
picture-changing (i.e.
$V$ has the same ghost-number as $ Z^n V$). Note that at zero picture, the
two definitions of ghost-number coincide.

\subsec {Scattering amplitudes}

Superstring scattering amplitudes are calculated by evaluating correlation
functions of BRST-invariant vertex operators on N=1 super-Riemann
surfaces. Although the details of multiloop amplitude calculations
are complicated, it will be useful to sketch the multiloop expression
before concentrating on the tree-level amplitude.
For the closed-string
scattering of $n$ states described by
the vertex operators $V_r$ ($r$=1 to $n$),
the RNS expression for the $g$-loop scattering amplitude 
is\ref\Verl
\ref\Verl{E. Verlinde and H. Verlinde,
Phys. Lett. B192 (1987) p.95.}
\eqn\amp{
\lambda^{2g-2}\sum_{I=1}^{2^{2g }} \prod_{i=1}^{3g-3+n} \int d^2 \tau_i
<|\xi(z_0)\prod_{j=1}^g \oint_{A_j} dz_j \eta(z_j) \delta(\oint_{A_j} dz_j
( \p\phi -i\xi\eta))}
$$\prod_{i=1}^{3g-3+n}\int d^2 u_i b(u_i) \mu_i(u_i) 
\prod_{s=1}^{2g-2+n-p} Z(y_s)|^2 \prod_{r=1}^n V_r(\tau_r)> $$
where 
$I$ labels the
$2^{2g}$ spin-structures for the worldsheet spinors (for each of the
$2g$ non-trivial loops on a genus $g$ surface, one can choose
periodicity or anti-periodicity conditions for the worldsheet spinors),
$\tau_i$ are Teichmuller parameters used
to describe the genus $g$ Riemann surface with $n$ punctures
(the first $n$ of these $3g-3+n$
parameters are the locations of the punctures),
$\mu_i$ are the corresponding Beltrami differentials, $\oint_{A_j}$ is
a contour integral around the $j^{th}$ $A$-cycle of the surface,
and the sum of the pictures of the external vertex operators is equal
to $p$.

The term 
\eqn\wata{
\xi(z_0)\prod_{j=1}^g \oint_{A_j} dz_j \eta(z_j) \delta(\oint_{A_j} dz_j
( \p\phi -i\xi\eta))}
comes from the need to restrict the picture of states propagating in
internal loops. 
\ref\watamura{U. Carow-Watamura, Z. Ezawa, K. Harada, A. Tezuka and 
S. Watamura, Phys. Lett. B227 (1989) p.73.}
Without this restriction, each propagating state would
be represented infinitely many times, violating unitarity. Although not
obvious, it can be checked that this term does not violate 
modular invariance (i.e. the choice of $A$ versus $B$-cycles is 
irrelevant). Note that the location of $z_0$ is arbitrary since only
the $\xi$ zero mode contributes.

The term
$\prod_{s=1}^{2g-2+n-p} Z(y_s)$ comes from integrating over the 
fermionic supermoduli of the N=1 super-Riemann surface where
the $y_s$ locations depend on the parameterization of the
supermoduli. One would expect that the scattering amplitude
should be independent
of the parameterization of the supermoduli, and therefore, independent
of the choice of the $y_s$ locations. Indeed, it is straightforward
to show that changing the
locations of the the $y_s$'s shifts the correlation function by
a total derivative in the $\tau_i$ parameters, which can be ignored
if the region of integration for the $\tau_i 's$ can be compactified
(i.e. if there are no singularities near the boundary of moduli space).\Verl

Note that the above expression for the scattering
amplitude is only spacetime-supersymmetric after summing over the
$2^{2g}$ spin-structures since, before summing, the correlation functions
contain unphysical
square-root cuts.
This makes it difficult to check finiteness
properties in the RNS formalism since, before summing over spin-structures,
the amplitude contains many unphysical divergences. As will be later shown,
the new description of the superstring does not suffer from this problem
since all fields are automatically GSO-projected, and there are therefore
no square-root cuts and no need to sum over spin structures. 

The tree-level scattering amplitude is much simpler
and is given by the expression
\eqn\tree{\lambda^{-2} 
< V_1(z_1) V_2(z_2) 
V_3(z_3) Z_L(z_3)Z_R(\bar z_3)}
$$
\prod_{r=4}^n\int d^2 z_r  [b_L, \{b_R , V_r(z_r)\}] Z_L(z_r) Z_R(
\bar z_r)>$$
where $\{b, V\}$ means the contour integral of $b$ around $V$, $V_r$
is in the $-1$ picture, and
the amplitude is independent of $z_1$, $z_2$, and $z_3$.

Despite the remarkable simplicity of this expression, 
it has two problems which
will be partially cured in the new superstring description.
Firstly, although the amplitude is spacetime-supersymmetric (i.e., 
$A(\{q_\a, V_1\}, ..., V_n)+ ... +$
$A(V_1, ..., \{q_\a, V_n\})=0$,
the spacetime-supersymmetry is far from manifest since Ramond
vertex operators look very different from Neveu-Schwarz
vertex operators. A second disadvantage
is that explicit calculations require SO(9,1)
Poincar\'e invariance to be broken to SU(5) (actually, a Wick rotated
version of SU(5)) in order that Ramond fields can be expressed in 
terms of the bosonized $\sigma_j$'s of \susy.

\subsec {Four-dimensional compactifications}

The N=1 c=15 superconformal representation with ten $x$'s and ten
$\psi$'s contains sixteen spacetime-supersymmetries, and is the most
symmetric representation for the superstring. However one can also
construct consistent superstring representations with fewer spacetime
supersymmetries.

One such representation is the ``four-dimensional'' superstring, which
contains four bosonic worldsheet scalars $x^m$ ($m$=0 to 3), four fermionic
worldsheet spinors $\psi^m_{L,R}$, and a c=9 N=2 superconformal field
theory which is described by the N=2 generators
$[T_{C},G^+_{C},G^-_{C}, J_{C}]$.
For different choices of the c=9 N=2 superconformal field theory, this
representation can be used to describe any four-dimensional compactification
of the ten-dimensional superstring which preserves at least N=1
4D spacetime-supersymmetry.\ref\gep{D. Gepner, Phys. Lett. B199 (1987)
p.380.}

The action and c=15 N=1 generators 
for this superconformal representation are: 
\eqn\comp{S={1\over 2\pi}\int d^2 z (\half \p x^m \bar\p x_m + i
\psi_L^m \bar\p\psi_{Lm}
+i\psi_R^m\p\psi_{Rm}) +S_{C}}
$$T=\half \p x^m \p x_m +{i\over 2} \psi^m \p \psi_m +T_{C}$$
$$G = \psi^m \p x_m + G^+_{C} +G^-_{C}$$
where $S_{C}$ is chosen such that $[T_{C},G_{C},\bar G_{C},
J_{C}]$ satisfy the following c=9 N=2 OPE's:
\eqn\twoOPE{G^+_C(y) G^-_C(z)\to {{3i}\over{(y-z)^3}}+
{{J_C(z)}\over{(y-z)^2}}+
{{iT_C+\half\p J_C}\over {y-z}},}
$$
T_C(y)T_C(z) \to {{9}\over{2(y-z)^4}}+{{2T_C}\over{(y-z)^2}}
+{{\p T_C}\over {y-z}}.$$

The simplest example of such an $S_C$ is
\eqn\simple{S_C={1\over 2\pi\a '}\int d^2 z ( \p x^j \bar\p \bar x_j + 
2i\psi_L^{j} \bar\p \bar\psi_{Lj}
+2i\psi_R^{j}\p\bar\psi_{Rj})}
$$T_C=\p x^j \p\bar x_j + 
{i\over 2}\psi^{j} \p \bar\psi_j+
{i\over 2}\bar\psi_{j} \p \psi^j, $$
$$G^+_{C} = \psi^{j} \p \bar x_j,\quad
G^-_{C} =\bar\psi_j \p  x^j,\quad
J_{C}= \psi^{j}\bar\psi_j,$$
where $j$=1 to 3. It is easy to check that the action and N=1 generators
of \act are the same as those of \comp where $x^j$ is identified with
$2^{-\half}(x^{3+j} +i x^{6+j})$, $\bar x_j$ is identified with   
$2^{-\half}(x^{3+j} -i x^{6+j})$, $\psi^{j}$ is identified with 
$2^{-\half}(\psi^{3+j} +i \psi^{6+j})$, and
$\bar\psi_j$ is identified with 
$2^{-\half}(\psi^{3+j} -i \psi^{6+j})$. 
This representation therefore corresponds
to an uncompactified superstring (if $x^j$ takes values in $R^6$)
or a toroidally-compactified superstring (if $x^j$ takes values in $T^6$).

As before, one can add an N=1 ghost system and construct a nilpotent
BRST operator. Physical vertex operators must satisfy the same conditions
as before, however there are now only four spacetime-supersymmetry
generators defined by: 
\eqn\foursusy{
q_a ={1\over{2\pi i}}
\oint dz e^{{i\over 2}(i\phi \pm \s_0 \pm \s_1\pm H_C)}}
where $\psi^3\pm\psi^0=e^{\pm i\s_0}, \psi^1\pm i \psi^2=e^{\pm i\s_1}$, 
$\p H_C= J_C$, and there are an even number of $+$ signs in the
exponent. Note that $H_C(y) H_C(z) \to -3 log (y-z)$ as $y\to z$, so
the integrand of $q_a$ has weight 1.\friedan

\newsec {The New Spacetime-Supersymmetric Description}

\subsec{Off-shell spacetime-supersymmetry generators}

In the RNS description, the spacetime-supersymmetry generators
of \susy satisfy the anti-commutation relations:
\eqn\alg{\{ q_a, q_b\}={1\over{2\pi i}}
\oint dz e^{-\phi} \psi_\m \gamma^\m_{ab}}
which is not the usual supersymmetry algebra
\eqn\usual{\{ q_a, q_b\}={1\over{2\pi }}
\oint dz \p x_\m \gamma^\m_{ab}}
(note that ${1\over 2\pi}\oint dz \p x_\m $ is the string momentum).

However after hitting the left-hand side of \alg with the picture-changing
operator $Z$, it becomes
$${1\over{2\pi i}}
\oint dz Z e^{-\phi} \psi_\m \gamma^\m_{ab}={1\over{2\pi }}
\oint dz 
\p x_\m \gamma^\m_{ab}.$$
So up to picture-changing, the $q_a$'s form a supersymmetry algebra.

But to make spacetime-supersymmetry manifest, one needs the $q_a$'s to
form a supersymmetry algebra without applying
picture-changing operations. This
is because manifest spacetime-supersymmetry requires
the generators to form an off-shell supersymmetry algebra, but
picture-changing is only well-defined when the
states are on-shell (otherwise, the states are not independent
of the locations of the picture-changing operators). 

So manifest spacetime-supersymmetry requires modification of the
$q_a$'s. Note that $q_a$ has picture $-\half$ and the momentum
${1\over{2\pi }}
\oint dz \p x_\m $
has picture 0, so we need generators with picture $+\half$.
The obvious guess is 
\eqn\guess{\bar q_a =Z q_a =
{1\over{2\pi i}}
\oint dz [b\eta e^{{i\over 2}(-3i\phi\pm\s_0\pm\s_1\pm\s_2\pm\s_3\pm\s_4)}}
$$
+i~:(e^{\phi}\psi_\m\p x^\m )
e^{{i \over 2}(i\phi\pm\s_0\pm\s_1\pm \s_2\pm\s_3\pm\s_4)}:].$$

It is easy to check that $\{q_a, \bar q_b\}=
{1\over{2\pi }}
\oint dz \p x_\m \gamma^\m_{ab}, $ so we now have a supersymmetry
algebra, but we also have twice too many supersymmetry generators! In ten
dimensions, it is not possible to keep half of the 32 generators in
an SO(9,1) Lorentz-covariant way. But for compactifications to
four or six
dimensions, it {\it is} possible to covariantly keep half of the generators.
Although we shall only discuss the four-dimensional
case in this review, the six-dimensional case has been discussed in 
reference \top. 

For compactifications to four dimensions, one can choose to keep the chiral
part of $q_a$ and the anti-chiral part of $\bar q_a$:\fourme
\eqn\sol{q_\a= 
{1\over{2\pi i}}
\oint dz  e^{{i\over 2}(i\phi\pm(\s_0 +\s_1)+ H_C)}}
$$\bar q_\ad=
{1\over{2\pi i}}
\oint dz [b\eta e^{{i\over 2}(-3i\phi\pm(\s_0-\s_1)- H_C)}$$
$$+i~:(e^{\phi}\psi_m\p x^m + 
e^{\phi}G_{C}^+ +
e^{\phi}G_{C}^-) e^{{i \over 2}(i\phi\pm(\s_0-\s_1)- H_C)}:].$$
These satisfy off-shell the 4D N=1 supersymmetry algebra
\eqn\fouralg{\{q_\a, \bar q_\bd\}=
{1\over{2\pi }}
\oint dz\p x_{\a\bd} }
where we are using the standard notation $x_{\a\bd}=x_m \s^m_{\a\bd}$.

\subsec {Hermiticity}

Although we have solved the problem of finding off-shell
super-Poincar\'e generators, we now have a new problem: 
using the standard RNS
definition of hermiticity where all fundamental fields are hermitian
or anti-hermitian (the anti-hermitian field is $\s_0$),
the hermitian conjugate of $q_\a$
is no longer $\bar q_\ad$. Fortunately, this new problem can
be solved by modifying the definition of hermiticity.\ref\hermit
{N. Berkovits, ``Off-shell supersymmetry versus hermiticity in the
superstring'', preprint IFUSP-P-1213, April 1996, to appear on hep-th.}

To find the appropriate hermiticity definition, one
first writes $\bar q_\ad$ of \sol in the form
\eqn\firstw{\bar q_\ad = e^R~({1\over{2\pi i}}
\oint dz b \eta e^{{i\over 2}
{(-3i\phi \pm (\s_0 -\s_1)-H)}}) ~e^{-R}}
where 
\eqn\Rdef{R={1\over{2\pi }}
\oint dz c \xi e^{-\phi} (\psi^m \p x_m +G^+_{C}+G^-_{C})}
and $e^R F e^{-R}= F +[R,F] +\half [R,[R,F]] + ...$ (the expansion usually
stops after two terms).
One can then define hermiticity as:
\eqn\herm{(x_m)^\dagger = e^R x_m e^{-R},\quad
(\psi_m)^\dagger = e^R \psi_m e^{-R},\quad
(F_C)^\dagger = e^R \bar F_C e^{-R},}
$$(e^{{\phi\over 2}})^\dagger = e^R (c  \xi e^{-{3\over 2}\phi}) e^{-R},\quad
(e^{-{\phi\over 2}})^\dagger = e^R (b  \eta e^{{3\over 2}\phi}) e^{-R},$$
$$(b)^\dagger = e^R (i b \eta \p\eta e^{2\phi}) e^{-R},\quad
(c)^\dagger = e^R (-i c \xi \p\xi e^{-2\phi}) e^{-R},$$
$$(\eta)^\dagger = e^R (i\eta b \p b e^{2\phi}) e^{-R},\quad
(\xi)^\dagger = e^R (-i\xi c \p c e^{-2\phi}) e^{-R},$$
where $F_C$ are the worldsheet fields in the c=9 N=2 superconformal
fields theory.

Since $R^\dagger= R$, 
\eqn\OK{(({F})^\dagger)^\dagger =(e^R \bar F e^{-R})^\dagger
=(e^{-R})^\dagger \bar F^\dagger (e^R)^\dagger =
e^{-R}(e^R F e^{-R}) e^R =F}
as desired.
It is straightforward to check that the new
hermiticity definition preserves all
OPE's
and implies that $(q_\a)^\dagger = \bar q_\ad $. 

One strange feature of the definition
is that a field may have a different conformal weight
from its hermitian conjugate since $(T)^\dagger= T+
\p (bc +\xi\eta)$. This will be explained in subsection 3.4 where
it will be related to the
``twisting'' of an N=2 superconformal field theory.

\subsec {Field redefinition}

The above definition of hermiticity appears to be much more complicated than
the standard RNS definition. 
However, one can now define a new set 
of worldsheet fields which are simple with respect to the new hermiticity
definition.\ref\redefme{
N. Berkovits, Nucl. Phys. B420 (1994) p.332.}\fourme
Besides simplifying the hermiticity definition, this new
set of fields will also allow spacetime-supersymmetry to be made manifest.

Since $q_\a$ is now the hermitian conjugate of $\bar q_\ad$, it is natural
to define fermionic superspace coordinates, $\t^\a$ and 
$(\t^\a)^\dagger = \tb^\ad$, which should satisfy
$\{q_\a, \t^\b\}=\delta_\a^\b$ and
$\{\bar q_\ad, \tb^\bd\}=\delta_\ad^\bd$. From the definitions of
$q_\a$ and $\bar q_\ad$, the natural candidates are
\eqn\cand{\t^\a = e^{{i\over 2}(-i\phi \pm (\s_0 + \s_1)-H)},\quad
\tb^\ad =  c \xi e^{{i\over 2}(3i\phi \pm (\s_0 - \s_1)+H)} .}

These fermionic coordinates have no singularities among themselves, and
are conjugate to
\eqn\conj{p_\a= e^{U} 
 e^{{i\over 2}(i\phi \pm (\s_0 + \s_1)+H)} e^{-U},}
$$
\bar p_\ad=(-p_\a)^\dagger =e^{U} 
( b \eta e^{{i\over 2}(-3i\phi \pm (\s_0 - \s_1)-H)}) e^{-U}$$
where 
\eqn\Udef{U = {1\over{2\pi}}
\oint dz c \xi e^{-\phi}( \half \psi_m \p x^m + G_C^-)}
(note that $U+U^\dagger = R$). 

In order that $x^m$ has no singularities with $p_\a$ or $\bar p_\ad$, we
shall redefine 
\eqn\redefine{x^m_{new}= e^U x^m_{old} e^{-U}
= x^m_{old}+ {i \over 2}
\theta^\a \s^m_{\a\ad}\tb_\ad.}
Note that $(x^m_{new})^\dagger= x^m_{new}$ with respect to the
new hermiticity definition. 

We shall also redefine the c=9 N=2 superconformal field theory so that
it has no singularities with $p_\a$ or $\bar p_\ad$. This is done by
redefining all the fields in the c=9 theory as
$$F_C^{new}= e^U (i\eta e^{\phi})^n F_C^{old} e^{-U}$$
where $n$ is the U(1) charge of $F_C^{old}$. (For example, for the
c=9 N=2 superconformal field theory of \simple, 
$x^j_{new} = x^j_{old}, ~\bar x_j^{new}=
\bar x_j^{old} +ic\xi e^{-\phi}\bar\psi_j^{old},~ 
\psi^j_{new} =i\eta e^{\phi} \psi^j_{old}+ ic\p x^j_{old},~
\bar \psi_j^{new}=
-i\xi e^{-\phi}\bar \psi_j^{old}.) $ 

This has the effect of redefining the c=9 N=2 generators to be:
\eqn\redefgen{T_C^{new} =e^U( T_C^{old} + {3\over 2} (\p\phi +i\eta\xi)^2 -
i(\p\phi +i\eta\xi) J_C^{old}) e^{-U},}
$$G_{C\,new}^+ = e^U (i \eta e^{\phi} G_{C\,old}^+) e^{-U},\quad
G_{C\,new}^- = e^U(-i\xi e^{-\phi} G_{C\,old}^-) e^{-U}
$$
$$J_C^{new}= J_C^{old} + 3i(\p\phi +i\eta\xi).$$
It is straightforward to check that these generators still form
a c=9 N=2 algebra with the standard hermiticity properties
(i.e. $(T_C^{new} )^\dagger =T_C^{new},~
(G_{C\,new}^+)^\dagger =G_{C\,new}^-,~
(J_C^{new})^\dagger =J_C^{new}$).

Besides the c=9 N=2 superconformal field theory, the original RNS system 
had five bosons ($x^m, \phi$) and eight fermions
($\psi^m, b,c,\eta,\xi$). Since
we so far have four bosons ($x^m$) and eight fermions ($\t^\a, \tb^\ad,
p_\a, \bar p_\ad$), there is one remaining boson which 
has no singularities in its OPE's with the
other fields. 
This boson, which will be called $\rho$, is 
chiral (like $\phi$) and is defined by:
\eqn\rhodef{\p\rho= -3i\p\phi+ cb + 2\xi\eta -J_C^{RNS}.}
Under hermitian conjugation, $(\rho)^\dagger =\rho$.

Besides simplifying the hermiticity properties, 
the new worldsheet
fields $[x_{new}^m, \t^\a,$
$\t^\ad, p_\a, \bar p_\ad,$
$\rho, F_C^{new}]$
have an important advantage over the original RNS fields.
Any operator constructed out of integer powers of the new fields
(or $e^{in\rho}$ where $n$ is an integer)
is automatically GSO-projected, i.e. it has no branch-cuts with the
spacetime-supersymmetry generators.
This property makes it possible to eliminate the sum over spin
structures, which is crucial for manifest spacetime-supersymmetry.

The RNS worldsheet action is simple to translate into the new fields
since the new fields have only free-field OPE's with each other. The
action is
\eqn\newac{
S={1\over 2\pi} \int d^2 z (\half\p x^m_{new} \bar\p x_m^{new} +
p_{L\a} \bar\p \t^\a_L +
\bar p_{L\ad} \bar\p \tb^\ad_L 
+p_{R\a} \p \t^\a_R +
+\bar p_{R\ad} \p \bar\t^\ad_R }
$$ +\half \p\rho_L\bar\p\rho_L
+\half \p\rho_R\bar\p\rho^R) +S_C^{new} $$
with free-field OPE's as $y\to z$:
\eqn\newope{
x^m (y)x^n(z)\to\eta^{mn} \log |y-z|^2,\quad \rho(y)\rho(z)
\to \log (y- z),}
$$p_\a(y) \t^\b(z)\to {{\d_\a^\b}\over {y -z}},\quad
\bar p_\ad(y) \tb^\bd(z)\to {{\d_\ad^\bd}\over {y -z}}.$$

For completeness,
we have temporarily unsuppressed the right-moving degrees of freedom in
the action.
Note that $\rho_L$ and $\rho_R$ are independent
left and right-moving chiral bosons. 

\subsec {Twisted N=2 structure}

Since $Q$ of equation \Q
is a useful operator in the RNS formalism, it is natural to
ask what is $(Q)^\dagger$? Writing 
\eqn\Qwrite{Q= {1\over{2\pi }}
\oint dz e^R (ib\eta\p\eta e^{2\phi}) e^{-R},}
it is easy to see that
$Q^\dagger ={1\over{2\pi }}
\oint dz b$. (In this way of writing $Q$, it is trivial to
check that $Q$ is nilpotent.)
If we define $j_{BRST}=e^R (ib\eta\p\eta e^{2\phi}) e^{-R}$, then
$(j_{BRST})^\dagger =b$. 

This hermiticity condition looks strange in an N=1 superconformal
system, but it is natural if we interpret
$j_{BRST}$ and $b$ as twisted fermionic
N=2 superconformal generators, $G^+$ and $G^-$. Such an interpretation
is possible since $j_{BRST}$ and $b$ satisfy the OPE's of 
twisted c=6 N=2 superconformal generators:\redefme\ref\ust
{N. Berkovits and C. Vafa, Mod. Phys. Lett. A9 (1994) p.653.}
\eqn\newtwo{G^+(y) G^-(z)\to {{2i}\over{(y-z)^3}}+
{{J(z)}\over{(y-z)^2}}+
{{iT}\over {y-z}},}
$$
T(y)T(z) \to {{2T}\over{(y-z)^2}}
+{{\p T}\over {y-z}}$$
where
\eqn\twist{T=T_{RNS},\quad
G^+= j_{BRST},\quad
G^-=b,\quad
J = bc+\eta\xi.}

In terms of the new fields defined in the previous subsection,
it is straightforward to calculate that
these twisted c=6 N=2 superconformal generators are:
\eqn\terms{T=T_{RNS}}
$$=\half \p x^m \p x_m +p_\a \p\t^\a +\bar p_\ad \p\tb^\ad
+\half\p\rho\p\rho +{i\over 2}\p^2\rho ~+T_C+{i\over 2} \p J_C,$$
$$G^+= j_{BRST}= e^{i\rho} d_\a d^\a ~ +G_C^+,$$
$$\bar G=b= e^{-i\rho} \bar d_\ad\bar d^\ad ~ + G_C^-,$$
$$J = bc+\eta\xi=-\p\rho ~+J_C,$$
where 
\eqn\ddef{d_\a=p_\a +{i\over 2} 
\tb^\ad \p x_{\a\ad}-{1\over 4}(\tb)^2 \p\t_\a +{1\over {8}}\t_\a\p(\tb)^2,}
$$\bar d_\ad=\bar p_\ad + {i\over 2}
\t^\a \p x_{\a\ad}-{1\over 4}(\t)^2 
\p\tb_\ad +{1\over {8}}\tb_\ad\p(\t)^2,$$
and $(\t)^2 =\half\epsilon_{\a\b}\t^\a\t^\b$, 
$(\tb)^2 =\half\epsilon_{\ad\bd}\t^\ad\t^\bd$. 
(In \terms and for the rest of this
paper, we will suppress the ``new'' label on the worldsheet
fields.) 

Similarly, the spacetime-supersymmetry generators of \sol can be calculated to
be
\eqn\susyterm{q_\a={1\over{2\pi i}}
\oint dz (p_\a - {i\over 2}
\tb^\ad \p x_{\a\ad}-{1\over {8}}\t_\a\p(\tb)^2),}
$$\bar q_\ad={1\over{2\pi i}}
\oint dz (\bar p_\ad - {i\over 2}
\tb^\ad \p x_{\a\ad}-{1\over {8}}\tb_\ad\p(\t)^2).$$
As was shown by Siegel,\ref\dSieg{
W. Siegel, Nucl. Phys. B263 (1986) p.93.}
$~d_\a$ and $\bar d_\ad$ anti-commute
with these spacetime-supersymmetry generators 
and satisfy the OPE that
$d_\a(y) d_\b(z)$ is regular,
\eqn\dope{d_\a(y) \bar d_\ad (z) \to 
{{i\Pi_{\a\ad}}\over {y-z}},\quad
d_\a(y) \Pi_{\b\bd}(z) = {-i\epsilon_{\a\b}\p\tb_\bd \over y-z}}
where $\Pi_{\a\ad}=\p x_{\a\ad}-{i\over 2}(\t_\a\p\tb_\ad +\tb_\ad\p\t_\a)$.

Since 
$T$ can be written as 
$$T=\half \Pi^m \Pi_m +d_\a \p\t^\a +\bar d_\ad \p\tb^\ad
+\half\p\rho\p\rho +{i\over 2}\p^2\rho ~+T_C+{i\over 2} \p J_C,$$
the above twisted c=6 N=2 generators are manifestly spacetime-supersymmetric
(note that four-dimensional
spacetime-supersymmetry now commutes with all 
compactification-dependent variables).
Another nice feature of these c=6 N=2 generators is that
they split cleanly into a set of
c=$-3$ N=2 generators (which depend on the ``four-dimensional'' fields)
and a set of c=9 N=2 generators (which depend on the compactification
fields).

So using RNS matter and ghost fields, we have constructed a set of
twisted c=6 N=2 generators which, when expressed in terms of new
variables, satisfy standard hermiticity properties and are manifestly
spacetime-supersymmetric.

Since c=6 is the critical central charge for an N=2 matter system, we can
now forget about its N=1 origin, untwist the N=2 generators by
shifting $T\to T-{i\over 2} J$,
introduce a set of c=$-6$ N=2 ghosts,
construct an N=2 BRST operator and vertex operators, and calculate
scattering amplitudes using standard N=2 techniques. This was the
method used in reference \ref\calcume{
N. Berkovits, Nucl. Phys. B395 (1993) p.77.}. Although it is
not obvious that the resulting N=2 prescription produces vertex
operators and scattering
amplitudes which coincide with those produced by the standard
N=1 RNS prescription, it was proven 
in reference \ust  that they
indeed do coincide.

However there is a simpler method to calculate scattering
amplitudes. Since the above N=2 matter system
was constructed out of RNS matter {\it and} ghost fields, there should be no
need to introduce additional N=2 ghosts. Note that $T(y)T(z)$ has
no central charge in the twisted
N=2 system, and all bosonic and fermionic worldsheet fields have integer
spin. So although there are certainly physical propagating states, the system
appears to be topological.

Indeed, scattering amplitudes for this system can be
calculated in a manner which is a direct N=4 generalization of 
N=2 topological techniques.\top
This topological method of calculation
can be generalized
to any c=6 N=2 system, and has been used to prove vanishing theorems
and calculate multiloop amplitudes for the c=6 N=2 system 
corresponding to 4D self-dual gravity.\top\ref\Oog{H. Ooguri and
C. Vafa, Nucl. Phys. B451 (1995) p.121.}

\newsec{N=4 topological method}

\subsec{N=4 generators}

To formulate the N=4 topological method, one first constructs  
a set of twisted c=6 N=4 generators out of the original twisted
c=6 N=2 generators.
These are defined by
\eqn\const{
T_{N=4}=T_{N=2}, \quad G^+_{N=4}=G^+_{N=2}, \quad G^-_{N=4}=G^-_{N=2},}
$$\tilde G^+_{N=4}=iG^-_{N=2}(e^{iH}),\quad 
\tilde G^-_{N=4}=-i G^+_{N=2}(e^{-iH}),$$
$$J_{N=4}= J_{N=2},\quad J^{++}_{N=4}= e^{iH},\quad
J^{--}_{N=4}=e^{-iH}$$
where $J_{N=2}=\p H$ and $G^+_{N=2} (e^{-iH})$ means the single pole in the
OPE of $e^{-iH}$ and $G^+_{N=2}$.

For a worldsheet primary field $f$ with U(1) charge $k$ (i.e.
$J (f)=ik f$), we shall define $\tilde f$ to be the pole of order $k^2$
in the OPE of $e^{ik H}$ and $\bar f$. Note that $\tilde f$ carries the same
U(1) charge and conformal weight as $f$, and $\tilde{\tilde f}=f$.

So 
$T$, $G^-$, $\tilde G^-$ and $J^{--}$ carry conformal weight 2,
$G^+$, $\tilde G^+$ and $J$ carry conformal weight 1, and
$J^{++}$ carries conformal weight 0.
It is straightforward to check that these generators satisfy the 
``small'' N=4 OPE's:
\eqn\fourope{G^+ \tilde G^- \sim G^- \tilde G^+ \sim 0}
$$G^+(z) G^-(0) \sim {{2i}\over {z^3}} +
{J(0)\over z^2} +{iT(0)\over z}$$
$$\tilde G^+(z) \tilde G^-(0) \sim -{{2i}\over{z^3}}
-{J(0)\over z^2} -{iT(0) 
\over z}$$
$$G^+(z)\tilde G^+(0) \sim {J^{++}(0)\over z^2}+
{\partial J^{++}(0)\over 2z}$$
$${G^-(z)\tilde G^-(0)\sim {J^{--}(0)\over z^2}+
{\partial J^{--}(0)\over 2z}}$$

In terms of the RNS and supersymmetric variables, the new generators are
given by
\eqn
\newtilde{
\tilde G^+= \eta =e^{i(-2\rho+H_C)} (\bar d)^2 +G^-_C(e^{i(-\rho+H_C)}),}
$$\tilde G^-= \{Q_{RNS},b\xi\}= b Z +\xi T_{RNS}
 =e^{i(2\rho-H_C)} (d)^2 +G^+_C(e^{i(\rho-H_C)}) ,$$
$$J^{++}= c \eta= e^{-i(\rho - H_C)},\quad J^{--}=b\xi =
e^{i(\rho - H_C)}.$$

\subsec{Physical vertex operators}

Since ${1\over{2\pi }}
\oint dz G^+= Q$ and ${1\over{2\pi }}
\oint dz \tilde G^+={1\over{2\pi }}
\oint dz \eta$, 
it is straightforward to translate the RNS language of section 2.2
into N=4 topological language. We shall find that by combining
N=4 topological language with the new supersymmetric variables, the
conditions on physical vertex operators can be greatly simplified.

Firstly, one needs to translate the requirement that the vertex operator
$V$ is in the BRST cohomology. This is simply
\eqn\req{V:\quad G^+(V)=\tilde G^+(V)=0,\quad V\neq G^+(\L)}
for any $\L$ satisfying $\tilde G^+(\L)=0.$ 

Note that $\tilde G^+(h)=0$ implies that $h=\tilde G^+ (f)$    
where $f=\xi h$. Similarly, $G^+(h)=0$ implies that
$h =G^+(f)$ where $f=c \xi\p\xi e^{-2\phi} h$. So 
$\oint dz G^+$ and $\oint dz \tilde G^+$ have trivial cohomologies.
Also note from \fourope that $G^+(y) \tilde G^+(z) \to 
(-\p_y +\p_z) (J(z)/2(y-z))$ as $y\to z$, so
$\{
\oint dz G^+,
\oint dz \tilde G^+\}=0$.

So the BRST cohomology condition of \req
can be written in a more symmetric form
as
\eqn\sym{
V: \quad G^+(V)= \tilde G^+(V)=0,\quad V\neq G^+(\tilde G^+(\Lambda))}
for {\it any} $\Lambda$.

Furthermore, since $\tilde G^+
(V)=0$ implies $V=\tilde G^+(\Phi)$ for $\Phi=\xi V$,
the BRST cohomology can also be defined as
\eqn\phidef{\Phi: \quad G^+(\tilde G^+
(\Phi))=0,\quad \Phi\neq G^+(\Lambda)+\tilde G^+(\tilde\Lambda).}
The $\tilde\Lambda$ gauge invariance comes from the ambiguity in
defining $\Phi$, since $\tilde G^+
(\Phi+ \tilde G^+(\tilde\Lambda))= \tilde G^+(\Phi)$.

We shall now translate the additional four physical conditions of section
2.2 into the new language:

1) The reality condition that $V=V^\dagger$ can not be directly
translated into the new language since $V^\dagger$ no longer has the
same conformal weight as $V$. Instead, the new reality condition will
be that $\Phi=\tilde\Phi$ where the $\tilde{}$ operation is defined 
below \const and
$V=G^+(\Phi)$.

2) The condition of being GSO-projected is trivially satisfied if the
vertex operator is a single-valued function of the new supersymmetric
variables. This is because integer powers of the worldsheet variables
are automatically GSO-projected. Note that $V$ is fermionic and
$\Phi$ is bosonic.

3) Although the picture-changing operator $Z$ does not play a role in
the N=4 topological method, there is a natural way to understand
picture-changing. As was just shown in \phidef, any BRST-invariant vertex
operator $V$ can be written as $V=\tilde G^+(\Phi)$ where
$G^+(\tilde G^+(\Phi))=0$.

Now consider the vertex operator $V^{(1)}= G^+(\Phi)$. Since
$\Gtp(G^+(\Phi))=G^+(\Gtp(\Phi))=0$, $G^+(V^{(1)})=\tilde G^+(V^{(1)})=0$, so
$V^{(1)}$ is also BRST-invariant. One can continue the procedure to obtain
$V^{(n)}$ for arbitrary positive $n$. It is easy to check that
$V^{(n)}= Z^n V + G^+(\Gtp(\Lambda^{(n)}))$ for some $\Lambda^{(n)}$.

Similarly, one can write $V=G^+(\Phi ')$ where
$\Gtp(G^+(\Phi '))=0$, and consider the operator $V^{(-1)}=\Gtp(\Phi ')$.
Once again, $V^{(-1)}$ is BRST-invariant, and the procedure can be continued
to give
$V^{(-n)}=Y^n V + G^+(\Gtp(\Lambda^{(-n)}))$ for some $\Lambda^{(-n)}$.

So for any N=4 topological theory, there is an infinite family of 
BRST-invariant vertex operators for each physical state, and one
has to choose a unique representative.\foot{For the N=4 topological
theory representing self-dual gravity, these vertex operators are related
by $V^{(n)}= e^{i\alpha n} V$ where $\alpha$ depends on the particle's
momentum.} In the RNS section, we learned that the picture eigenvalue $p$
could be used to select a unique representative, but this breaks manifest
spacetime-supersymmetry since $[q_\a, P]\neq 0$.

Writing the picture operator $P$ in terms of the new variables,
$$P={1\over{2\pi }}
\oint dz(i\xi\eta-\p\phi) ={1\over{2\pi }}
\oint dz (i\p\rho -\half p_\a \t^\a +
\half\bar p_\ad \tb^\ad),$$
it is obvious that $[P,q_\a]\neq 0$ since $P$ contains $\t^\a$
zero modes.
However, if we use $\rho$ charge instead of picture to select
a unique representative, spacetime-supersymmetry is preserved since
$ \oint dz \p\rho$ commutes with $q_\a$ and $\bar q_\ad$.

As elaborated in reference \ref\fieldme{
N. Berkovits, Nucl. Phys. B450 (1995) p.90.}, it is possible to restrict all 
compactification-independent states to have zero $\rho$-charge,
and all compactification-dependent states to have $-1$, 0, or $+1$
$\rho$-charge. It was proven in reference \fieldme
that this restriction selects
a unique representative for each physical state in a manner that
preserves manifest SO(3,1) super-Poincar\'e invariance.

4) The final condition of $+1$ ghost-number is easily translated into
N=4 topological language since the ghost-number current,
\eqn\newgh{J_g= {i\over{2\pi }}
\oint dz (bc+\xi\eta)={1\over{2\pi }}
\oint dz (-\p\rho + J_C),}
is also the U(1) current. So $V$ must have $+1$ U(1)-charge, which
means that $\Phi$ is U(1)-neutral.
Note that $\Phi=\tilde\Phi$ implies $\Phi = \Phi^\dagger$ when
$\Phi$ is U(1)-neutral.

So for compactification-independent states, $\Phi$ uniquely represents
a physical state if $\Phi$ is a real single-valued bosonic function of
four-dimensional fields, is U(1)-neutral, satisfies
$G^+(\Gtp(\Phi))=0$, and cannot be written as $\Phi=G(\L) + 
\Gtp(\tilde \L)$ for
any $\L$ and $\tilde \L$. Note that U(1)-neutrality implies zero $\rho$-charge
since $\Phi$ is independent of the compactification fields. (For
simplicity, only compactification-independent states will be discussed in
this
review. 
Details on compactification-dependent states
can be found in references \fourme, \fieldme and 
\ref\mesieg
{N. Berkovits and W. Siegel, ``Superspace effective actions for
4D compactifications of heterotic and type II superstrings'',
preprint USPIF-P-1180, October 1995, to appear in Nucl. Phys. B,
hep-th 9510106.}).

This N=4 topological definition of physical vertex operators was obtained
by comparing with the N=1 RNS definition, and naively appears to be 
unrelated to the standard definition coming from a critical N=2
superconformal field theory. In critical N=2 superconformal field theories,
physical vertex operators correspond to bosonic
primary fields $\Phi$ of zero conformal weight and zero U(1)-charge (i.e.,
$\Phi$ has no double poles
with the N=2 generators $T$, $G^+$, and $G^-$,
and has no single pole with $J$).
In integrated form, the vertex operator for open N=2 strings is given by
$$W=\int dz G^- (G^+ (\Phi))$$
where $G^-( X )$ means the single pole in the OPE of $G^-$ and 
$X$.\foot{
For the N=2 open string representing 4D
self-dual Yang-Mills, $\Phi(x)$ is the
Yang scalar and $W=\int dz ( \p x^j \p_j\Phi +i\psi^j\bar\psi_k \p_j\bar \p^k
\Phi)$.}

Although not obvious, it was proven in references \ust and \ref\ohta
{N. Ohta and J.L. Petersen, Phys. Lett. B325 (1994) p.67.} that the 
N=1 and N=2 definitions of BRST cohomology coicide, and there is therefore
a one-to-one correspondence between $\Phi$'s satisfying the N=4
topological definition and $\Phi$'s satisfying the standard N=2 definition.
Note that 
\eqn\note{W=\int dz b(V)=\int dz G^-(V) =\int dz G^-(G^+(\Phi))}
is the integrated form of the RNS vertex operator, so the integrated
vertex operators also agree in the two definitions.

As an example, we shall now describe the vertex operators for all massless
states of the open superstring which are independent of the
compactification.
It will be seen that the new spacetime-supersymmetric variables allow these
vertex operators to be expressed in much more compact form than when using
the RNS variables.

\subsec{Massless compactification-independent states}

For the open superstring, the massless
compactification-independent states
are those of 4D N=1 super-Yang-Mills, where the gauge group comes from
Chan-Paton factors. The vertex operator $\Phi$ for these fields only
depends on the $[x^m,\t^\a,\tb^\ad]$ zero modes and is simply
the super-Yang-Mills prepotential
$\Phi^I (x,\t,\tb)$, which contains the gluon field in its $\t\tb$
component and the gluino fields in its $(\t)^2\tb$ and $\t(\tb)^2$ 
components ($I$ labels the gauge group generators).

Using the N=4 topological definition of \phidef, the on-shell condition is
\eqn\onshell{
\Gtp(G^+(\Phi^I)= e^{iH_C -i \rho} \bar D_\ad (D)^2 \bar D^\ad \Phi^I =0,}
and the gauge-invariances are 
\eqn\gauge{\d \Phi^I=G^+(\L^I) +\Gtp(\tilde\L^I) = (D)^2 \lambda^I + 
(\bar D)^2\bar\lambda^I}
where $\L^I= e^{-i\rho}\lambda^I$ and $\tilde\L^I =
e^{-iH_C+2i\rho}\bar\lambda^I$.
These imply the 
usual linearized equations of motion and gauge-invariances for the
super-Yang-Mills prepotential.

Using the standard N=2 definition, $\Phi^I$ is a physical field if it
has no double poles with $G^+$, $G^-$, or $T$, i.e. if $\Phi^I$ satisfies
the conditions 
$$(D)^2 \Phi^I = (\bar D)^2 \Phi^I=\p_m\p^m \Phi^I=0.$$
Note that $(D)^2 \Phi^I
=(\bar D)^2 \Phi^I=0$ is the supersymmetric version of the
Lorentz gauge condition $\p^m A^{I}_m =0$, and in this gauge, 
$\p^m \p_m \Phi^I=0$
is the linearized equation of motion.

In integrated form, the open superstring vertex operator is
\eqn\integ{W= 
\int dz
(d^\a ~(\bar D)^2 D_\a \Phi^I +
\bar d^\ad ~(D)^2\bar D_\ad \Phi^I }
$$+
\p\t^\a D_\a \Phi^I -\dzm\tba~ \bar D_\ad \Phi^I -i
\Pi^{\a\ad}~[D_\a,\bar D_\ad] \Phi^I ).$$

\subsec {Scattering amplitudes}

One way to compute scattering amplitudes in the new supersymmetric
description is using standard N=2 techniques, where one
integrates correlation functions of
BRST-invariant vertex operators on N=2 super-Riemann
surfaces. Although this N=2 prescription was successfully used in reference
\calcume,
it was later shown in reference \top that the N=4 topological method
provides a simpler prescription and produces equivalent amplitudes.
For this reason, only the N=4 topological prescription will be
described in this review.
As in section 2.3, we shall begin by briefly describing multiloop
amplitudes, and then concentrate on the simpler case of tree amplitudes.

The simplest derivation of the
N=4 topological prescription is to start with the N=1 prescription
of \amp, and translate it into N=4 topological language.
This is done in three steps:

1) Choose the $2g-2+n-p$ picture-changing operators
to be located at $2g-2+n-p$ of the $3g-3+n$ $b$ insertions.
2) Choose the $\xi(z_0)$ insertion to be located at one of the remaining
$b$ insertions.
3) Change the picture restriction, $\delta(\oint_{A_j} dz_j(
\p\phi-i\xi\eta))$, to a restriction on the $\rho$ charge,
$\delta(\oint_{A_j} dz_j \p\rho)$.

The first two steps are allowed since the amplitude is independent of the
locations of the picture-changing operators and of the $\xi$ zero mode.
The third step is allowed since, just like picture, the $\rho$ charge
can be used to select a unique representative for propagating states.

The RNS expression of \amp can now be written as:
\eqn\newamp{
\lambda^{2g-2}\sum_{I=1}^{2^{2g }} \prod_{i=1}^{3g-3+n} \int d^2 \tau_i
<|\prod_{j=1}^g \oint_{A_j} dz_j \eta(z_j) \delta(\oint_{A_j} dz_j
( \p\rho))}
$$\int d^2 u_1 \xi b(u_1) \mu_1(u_1) 
\prod_{i=2}^{g-1+p}\int d^2 u_i  b(u_i) \mu_i(u_i) 
\prod_{j=g+p}^{3g-3+n}\int d^2 u_j Z b(u_j) \mu_j(u_j)|^2 $$
$$
\prod_{r=1}^n [{1\over{2\pi}}\int d\bar z~\eta_R,~\{{1\over {2\pi}}\int dz
~\eta_L,~\xi V_r(\tau_r)\}]> $$
which easily translates into N=4 topological language as:
\foot{
After removing the 
$\delta(\oint_{A_j} dz_j \p\rho)$ term, the topological
prescription can also be used to compute
multiloop amplitudes for the N=2 string representing N=(2,2)
self-dual gravity.\top\Oog}
\eqn\trans
{\lambda^{2g-2}\sum_{I=1}^{2^{2g }} \prod_{i=1}^{3g-3+n} \int d^2 \tau_i
<|\prod_{j=1}^g \oint_{A_j} dz_j \tilde G^+(z_j) \delta(\oint_{A_j} dz_j
( \p\rho))}
$$\int d^2 u_1 J^{--}(u_1) \mu_1(u_1) 
\prod_{i=2}^{g-1+p}\int d^2 u_i  G^-(u_i) \mu_i(u_i) 
\prod_{j=g+p}^{3g-3+n}\int d^2 u_j \tilde G^-(u_j) \mu_j(u_j) |^2$$
$$
\prod_{r=1}^n  G^+_R(G^+_L(\Phi_r(\tau_r))>.
$$
Note that using the new spacetime-supersymmetric variables, there
is no need to perform a sum over spin structures. 

For tree-level amplitudes, \trans reduces for open strings to: 
\eqn\newtree{
< \Phi_1(z_1) \tilde G^+(\Phi_2(z_2)) 
G^+(\Phi_3(z_3))
\prod_{r=4}^n\int d z_r  G^-(G^+(\Phi_r(z_r))) .}

Plugging in the expression in section 4.3 for $\Phi$, it is
completely straightforward to compute massless
tree amplitudes in a manifestly
SO(3,1) super-Poincar\'e invariant manner. For example, the amplitude
for three super-Yang-Mills particles is:\foot{I would like to thank
Konstantin Bobkov for pointing out an error in the original version
of the three-point amplitude.}
\eqn\threeYM{
f_{IJK}<\Phi^I_1(z_1) \tilde G^+(\Phi^J_2(z_2)) G^+(\Phi^K_3(z_3))>}
$$
= f_{IJK}<\Phi^I_1(z_1)~ e^{-2i\rho+iH_C}\bar d^\ad(z_2)
\bar D_\ad \Phi^J_2(z_2)~ e^{i\rho}d^\a(z_3) D_\a\Phi^K_3(z_3)>$$
$$= f_{IJK}\int d^4 x\int d^2\t d^2\tb ~\Phi^I_1~ 
(D^\a \bar D_\ad \Phi^J_2~ \bar D^\ad D_\a\Phi^K_3
+ (k_2^{\a\ad}-k_3^{\a\ad})
\bar D_\ad \Phi^J_2~ D_\a\Phi^K_3 )$$
where $f_{IJK}$ is the structure constant and
$\int d^2\t d^2\tb$ comes from the background
charge condition
$<(\t)^2(\tb)^2 e^{-i\rho+iH_C}>=1.$

\newsec{Relationship with the Green-Schwarz formalism}

In this section, we shall review the light-cone,
conventional, and twistor-like
Green-Schwarz descriptions of the superstring, and then show the
relationship with the new description. 

\subsec{Light-cone Green-Schwarz formalism}

All Green-Schwarz descriptions of the superstring reduce in
light-cone gauge to the free-field action:\ref\schw{
J.H. Schwarz, Phys. Rep. 89 (1989) p.223.}
\eqn\LCGS{ {1\over{2\pi}}
\int d^2 z (\half\p x^i \bar \p x^i +i\t^a_L\bar\p \t^a_L
+i\t^a_R\p \t^a_R)}
where  $x^i$ is an SO(D-2) vector and $\t^a$ is an SO(D-2) spinor.
This light-cone gauge is defined by 
\eqn\LCdef{x^0+x^{D-1}=\tau,\quad
\p(x^0 -x^{D-1}) =\p x^i \p x^i + i\t^a \p \t^a,\quad
(\g^0 + \g^{D-1})_{\a\b} \t^\a =0 }
where $\tau$ is the worldsheet time, $x^\mu$
is an SO(1,D-1) vector, $\t^\a$ is an SO(1,D-1) spinor,
and 
$(\g^0 - \g^{D-1})_{\b a} \t^\b =\t_a$ is the light-cone spinor.

Although scattering amplitudes can be computed in light-cone gauge,
the computations are complicated by the fact that light-cone diagrams
have singular interaction points.\ref\Green
{M.B. Green and J.H. Schwarz, Nucl. Phys. B243 (1984) p.475.}
\ref\Mandelstam{S. Mandelstam, Prog. Theor. Phys. Suppl. 86 (1986)
p.163.}
\ref\Taylor{A. Restuccia and J.G. Taylor,
Phys. Rep. 174 (1989) p.283.}
At these interaction points, non-trivial
operators need to be inserted in order to preserve Lorentz invariance.
(These insertions are also necessary in the light-cone RNS description, but
not in the light-cone description of the bosonic superstring.\ref\lcM
{S. Mandelstam, Nucl. Phys. B69 (1974) p.77.}) The
non-trivial operators make it extremely difficult to write the amplitude
in closed form, and for this reason, only four-point tree and one-loop
superstring amplitudes have been computed using this light-cone method.\foot{
Although reference \Taylor
contains explicit expressions for multiloop amplitudes,
these expressions contain unphysical divergences when 
interaction points coincide. It has not yet been determined
how these expressions are affected by removing the
unphysical divergences with contact terms.
Also, Mandelstam has constructed an N-point
tree amplitude which is spacetime-supersymmetric,
but his proposal is based on unitarity arguments, rather than
explicit light-cone
calculations.\ref\Mandtwo{S. Mandelstam, {\it Workshop on
Unified String Theories, 29 July - 16 August 1985}, eds.
M. Green and D. Gross, World Scientific, Singapore (1986) p.577.}}

\subsec{Conventional Green-Schwarz formalism}

Based on earlier work by Siegel on the superparticle,\ref\siegpart{W. Siegel,
Phys. Lett. B128 (1983) p.397.}
Green and Schwarz found a super-Poincar\'e invariant action for the
superstring (in $D$=3,4,6 or 10) which reduces
to \LCGS in light-cone gauge.\ref\gstwo{M.B. Green and J.H.Schwarz,
Nucl. Phys. B243 (1984) p.285.}
In worldsheet conformal gauge, this action is
\eqn\GSact{
{\cal S}={1\over{2\pi}}\int d^2 z [\Pi^\mu\bar\Pi_{\mu} +
{i\over 2}(\p x_\m +
{i\over 2}
\t_L\g_\m\p\t_L) (\t_L\g^\m\bar\p\t_L - \t_R\g^\m \bar\p\t_R)}
$$-{i\over 2}
(\bar\p x_\m+{i\over 2}
\t_L\g_\m\bar\p\t_L) (\t_L\g^\m\p\t_L - \t_R\g^\m \p\t_R)]$$
with Virasoro constraint $T=\Pi^\mu \Pi_\mu$
where 
$$\Pi^\mu=\p x^\m -{i\over 2}
\t_L\g^m\p\t_L -{i\over 2}
\t_R\g^m\p\t_R.$$

Since the conjugate momentum for $\t^\a$ is 
$P_{\t^\a}=\p {\cal S}/\p(\p_0\t^\a)$, there is a spinor
Dirac constraint:
\eqn\second{d_\a=P_{\t^\a}+{i\over 2}\Pi_\m \g^\m_{\a\b} \t^\b=0.}
It is easy to compute that $\{d_\a,d_\b\}=i\Pi_\m \g^\m_{\a\b}$,
and since $\Pi_\m \Pi^\m$ is the Virasoro constraint, half of the
$2D-4$ components of $d_\a$ are first-class constraints and the other
half are second-class.
The first-class constraints in $d^\a$ generate the $D-2$ fermionic
$\k$-symmetries:
\eqn\kaPPa{\d\t^\a =\Pi^\m \g_\m^{\a\b} \k_\b,\quad
\d x^\m ={i\over 2}\t\g^\m\d\t,}
which allows half of the $\t$'s to be gauge-fixed.
(Note that the Virasoro constraint $\Pi_\m \Pi^\m$ implies
that only half of the $2D-4$ components of $\k_\b$ contribute to
$\d\t^\a$.)

Since \GSact is not a quadratic action, quantization is not straightforward.
Although \GSact simplifies somewhat in the ``semi-light-cone gauge''
$(\g^0 + \g^{D-1})_{\a\b} \t^\a =0$ to\ref\Car{S. Carlip,
Nucl. Phys. B284 (1987) p.365.}
\eqn\semilc{{1\over 2\pi}\int d^2 z (\half\p x^\m \bar\p x_\m +}
$$ 
i[\t_{L} (\g^0 -\g^{D-1})\bar\p
\t_L ]\p (x^0 + x^{D-1})
+i[\t_{R} (\g^0 -\g^{D-1})\p
\t_R] \bar\p (x^0 + x^{D-1})),$$
even the semi-light-cone gauge action is difficult to quantize.\ref\RK
{U. Kraemmer and A. Rebhan, Phys. Lett. B236 (1990) p.255.} 
Because
of these quantization problems, neither \GSact nor \semilc
has been successfully
used to compute superstring scattering amplitudes.

\subsec{Twistor-like Green-Schwarz formalism}

In \ref\STVZ
{D. Sorokin, V.I. Tkach, D.V. Volkov and A.A. Zheltukhin,
Phys. Lett. B216 (1989) p. 302.}, 
Sorokin, Tkach, Volkov, and Zheltukhin discovered an alternative
super-Poincar\'e invariant action which also reduces to \LCGS in 
light-cone gauge.\foot{Actually, reference \STVZ only discusses the
superparticle, but their work was soon generalized by other authors
to the heterotic superstring.\ref\twisthet
{N. Berkovits, Phys. Lett. B232 (1989) p.184\semi
M. Tonin, Phys. Lett. B266 (1991) p.312\semi
E.A. Ivanov and A.A. Kapustnikov, Phys. Lett. B267 (1991)
p.175\semi
F. Delduc, A. Galperin, P. Howe and
E. Sokatchev, Phys. Rev. D47 (1993) p.578.}
A simple twistor-like action for the Type II
superstring is still lacking.}
These authors discovered that the fermionic $\kappa$-symmetries of \kaPPa
could be converted into superconformal invariances if one introduced
bosonic spacetime spinor variables into the superstring action. These
new bosonic variables, $\lambda^\a$, are the worldsheet supersymmetric
partners of $\t^\a$.

In order to preserve the number of physical degrees of freedom, 
$\lambda^\a$ must be constrained to satisfy 
\eqn\twistor{\lambda^\a \g^\mu_{\a\b}
\lambda^\b = \Pi^\mu.}
Since the zero mode of \twistor resembles the twistor relation of Penrose,
\ref\Pen{R. Penrose and M.H. MacCallum, Phys. Rep. 6 (1972) p.241\semi
A. Ferber, Nucl. Phys. B132 (1978) p.55\semi
E. Witten, Nucl. Phys. B266 (1986) p.245\semi
I. Bengtsson and M. Cederwall, Nucl. Phys. B302 (1988) p.81.}
$\lambda^\a \g^\mu_{\a\b}
\lambda^\b = P^\mu$ where $P^\mu$ is the particle momentum,
the resulting string is called the twistor-like
Green-Schwarz superstring.

There is also a fermionic worldsheet superpartner of \twistor,
\eqn\twistorferm{
\lambda^\a \g^\mu_{\a\b}
\t^\b = \psi^\mu,}
which relates the RNS fermionic vector $\psi^\mu$ with the
GS fermionic spinor $\t^\a$. By combining the component fields
into the worldsheet superfields,
\eqn\super{X^\mu = x^\mu +i\kappa \psi^\mu,\quad
\Theta^\a = \t^\a + \k\lambda^\a,}
the constraints of \twistor and
\twistorferm can be expressed in worldsheet superconformally invariant
notation as
\eqn\superconst{
D_\kappa X^\mu =i \Theta^\a \g^\mu_{\a\b} D_\kappa \Theta^\b,}
where $D_\kappa= \p_\kappa+i \kappa\p_z$.

Although the invariances of the twistor-like approach are more
geometrical than the $\kappa$-symmetries of the conventional approach,
the twistor-like action is equally difficult to quantize. In $D=3,4,6$
or 10, one has up to $D-2$ superconformal invariances (which replace
up to $D-2$ $\kappa$-symmetries). So when $D=10$,
one can have N=8 super-Virasoro, which is not straightforward to 
quantize since the algebra is soft.\ref\soft{F. Englert, A. Sevrin,
W. Troost, A. Van Proeyen and P. Spindel, J. Math. Phys. 29 (1988) p.281\semi
N. Berkovits, Phys. Lett. B241 (1990) p.497\semi L. Brink, M. Cederwall
and C. Preitschopf, Phys. Lett. B311 (1993) p.76\semi M. Cederwall and
C. Preitschopf, Comm. Math. Phys. 167 (1995) p.373\semi
J.A. Henning Samtleben, Nucl. Phys. B453 (1995) p.429.}
Another problem is that
in addition to the fermionic second-class constraints of \second, one 
has bosonic second-class constraints coming from \twistor.\ref\problemme
{N. Berkovits, Nucl. Phys. B358 (1991) p.169.}

\subsec {Relationship with the new description}

Because the twistor-like approach contains worldsheet superconformal
invariance, one would suspect it is closely related to the new
formalism described in the previous sections. Indeed, it was shown in 
\ref\meworld{N. Berkovits, Nucl. Phys. B379 (1992) p.96.}
that after partially gauge-fixing the $D=10$ 
twistor-like action (which fixes
six of the eight fermionic worldsheet invariances and breaks
SO(1,9) super-Poincar\'e invariance down to SU(4)$\times$U(1)),
one can obtain a free-field action with critical N=2 superconformal invariance.
After a series of complicated field-redefinitions, this N=2 action can
be written as \newac, which was how \newac was originally found.

After gauge-fixing,
four components of 
the twistor-like variables are related to the worldsheet
variables of the previous sections by
\eqn\twistorrelat{\lambda^\a =d^\a e^{i\rho},\quad
\bar\lambda^\ad=\bar d^\ad e^{-i\rho}.}
Note that the OPE's of
$d^\a$ and $\rho$ imply that
$$\lambda^\a(y)\bar\lambda^\ad(z) \to
\Pi^m \s_m^{\a\ad}$$
as $y\to z$, which is the twistor constraint of \twistor. Also note that
the four-dimensional part of the N=2 fermionic generators in \terms are
\eqn\alsonote{
G^+ =d_\a d^\a e^{i\rho} = d_\a \lambda^\a,\quad
G^- =\bar d_\ad \bar d^\ad e^{-i\rho} = \bar d_\ad\bar\lambda^\ad,}
which implies that
$\lambda^\a$ and $\bar\lambda^\ad$ are the worldsheet superpartners of
$\t^\a$ and $\bar\t^\ad$.

Finally, we shall show that in light-cone gauge, the action of \newac
reduces to the standard light-cone GS action of \LCGS. 
This proves the classical
equivalence of the new description with the conventional and
twistor-like GS descriptions.

The first step is to fermionize one component of $\t^\a$ and $p_\a$ as
\eqn\lccheck{\t^1 = e^{i\s},\quad p_1 = e^{-i\s}}
where $\s$ is a chiral boson. The second step is to use the U(1) invariance
generated by $J=-\p\rho + J_C$ to gauge-fix $\rho=\s$.
The third step is to use the Virasoro constraint $T$ to gauge-fix
$x^0+ x^3 =\tau$.

At this point, the fermionic N=2 generators are
\eqn\now{
G^+ = d_\a d^\a e^{i\rho} + G^+_C= p_1 p_2 e^{i\rho} + ... = p_2 + ...,}
$$G^- = \bar d_\ad\bar d^\ad e^{-i\rho} + G^-_C=
\p( x^0+x^3)\t^1\bar p_2 e^{-i\rho} 
+ ... = \bar p_2 + ...,$$
where the $\p(x^0+x^3)\t^1$ term comes from $\bar d_1$.

The next step is to use $G^+$ and $G^-$ to gauge-fix $\t^2=\bar\t^2=0$.
Besides the compactification-dependent fields, this leaves
$[x^0 - x^3,x^1,x^2,$
$\bar\t^1,\bar p_1, p_2, \bar p_2,\rho]$.
However $x^0 - x^3$, $p_2,$ $\bar p_2,$ and $\rho$ are constrained by
$T=G^+=G^-=J=0$.

So the only physical unconstrained variables are $x^1$, 
$x^2$, 
$\bar \t^1$,
$\bar p_1$, 
and the compactification dependent fields. In this gauge, the
action for these fields is
\eqn\thisgauge{{1\over 2\pi}\int d^2 z (\half \p x^1 \bar\p x^1
+ \half\p x^2 \bar\p x^2 +\bar p_{L1}\bar\p\bar\t^1_L
+ \bar p_{R1}\p\bar \t^1_R) + S_C
,}
which is equal to \LCGS if the compactification is flat and
$[\bar\t^1,\bar p_1,\psi^{j},\bar\psi_j]$ are identified with the
eight light-cone $\t$'s ($\psi^{j}$ and $\bar\psi_j$ are defined in
\simple where $j=1$ to 3).

\newsec{Conclusions and Applications}

In this review, we have introduced a new spacetime-supersymmetric 
description of the superstring which has N=2 worldsheet superconformal
invariance. 
This description
is manifestly SO(3,1) super-Poincar\'e invariant for
arbitrary compactifications to four dimensions which
preserve N=1 4D supersymmetry. It is related to
the N=1 RNS description by embedding in an N=2 string and performing
a field redefinition. It is related to the twistor-like GS description
by gauge-fixing six of the eight fermionic worldsheet invariances.

There are three types of applications which have been developed for this
new superstring description. One application is the explicit 
spacetime-supersymmetric computation of superstring scattering amplitudes.
For N-point tree amplitudes, Koba-Nielsen formulas have been computed
which are manifestly SO(3,1) super-Poincar\'e invariant.\ref\koba
{N. Berkovits, ``Super-Poincar\'e invariant Koba-Nielsen formulas for the
superstring'', preprint IFUSP-P-1211, April 1996, to appear on hep-th.}
These formulas are new and can be generalized to scattering in the
presence of a $D$-brane.
For certain multiloop amplitudes involving Ramond-Ramond states, explicit
``topological'' expressions have been computed to all
loop-order for compactifications to
four and six dimensions.\top The four-dimensional expressions reproduce
the topological results of \ref\Antoniadis{I. Antoniadis, E. Gava, 
K.S. Narain and T.R. Taylor, Nucl. Phys. B413 (1994) p.162.}, while
the six-dimensional expressions are new.

It would be nice to have explicit spacetime-supersymmetric
expressions for arbitrary multiloop
amplitudes, and not just for ``topological'' ones. 
These expressions could be used for analyzing finiteness properties, which
is difficult in the RNS formalism because of the need to sum over
spin structures. The only obstacle to calculating multiloop amplitudes in the
new description is evaluating the correlation function of the chiral boson
$\rho$, which may
have unphysical poles (these
unphysical poles are absent in topological
amplitudes). Unphysical poles also occur for the
RNS chiral boson $\phi$, and hopefully, this
obstacle can be overcome using methods
similar to those of references \Verl and \watamura. 

A second type of application has been the construction of 
spacetime supersymmetric sigma models, which can be used to derive
the low-energy equations of motion in superspace for the
massless superstring fields. Unlike the standard
GS sigma model,\ref\wit{E. Witten, Nucl. Phys. B266 (1986) p.245.}
these sigma models contain a Fradkin-Tseytlin term
which couples the spacetime dilaton to the worldsheet (super)curvature.\ref
\mesig{N. Berkovits, Phys. Lett. B304 (1993) p.249.}
\mesieg

For 4D compactifications of the heterotic superstring, it has been
verified to one-loop order in $\a '$ that worldsheet N=(2,0) superconformal
invariance of the sigma model implies the standard superspace equations of
motion for the N=1 supergravity superfields.\ref\boer{J. de Boer
and K. Skenderis, private communication.} For 4D compactifications of the
type II superstring, properties of the sigma model have been used to obtain
new superspace actions for N=2 supergravity.\mesieg 
It would be interesting to
use worldsheet N=(2,2) superconformal invariance of the type II sigma model to
check if the low-energy superstring equations of motion come from these
new N=2 supergravity actions.

A third type of application has been the construction of an open superstring
field theory action.\fieldme
Unlike the RNS field theory action,\ref\witfield
{E. Witten, Nucl. Phys. B276 (1986), p.291.}, this new action is 
manifestly SO(3,1) super-Poincar\'e invariant and does not suffer from
contact-term divergences. Work is in progress on generalizing the construction
for closed superstring field theory. Such an action might be very useful
for studying perturbative and non-perturbative duality symmetries of the
superstring.
\listrefs
\end